%% file: FGCNN_for_www2018.tex
\documentclass[sigconf]{acmart}

\usepackage{booktabs} 
\usepackage{float}
\usepackage[normalem]{ulem}
\usepackage{graphicx}
\usepackage{footnote}
\makesavenoteenv{tabular}
\usepackage{booktabs}
\makesavenoteenv{table}
\usepackage{setspace}
\usepackage{subfigure}
\useunder{\uline}{\ul}{}
\usepackage{balance} 
\newenvironment{shrinkeq}[1]
{ \bgroup
  \addtolength\abovedisplayshortskip{#1}
  \addtolength\abovedisplayskip{#1}
  \addtolength\belowdisplayshortskip{#1}
  \addtolength\belowdisplayskip{#1}}
{\egroup\ignorespacesafterend}

\copyrightyear{2019}
\acmYear{2019}
\setcopyright{iw3c2w3}
\acmConference[WWW '19]{Proceedings of the 2019 World Wide Web Conference}{May 13--17,
2019}{San Francisco, CA, USA}
\acmBooktitle{Proceedings of the 2019 World Wide Web Conference (WWW '19), May 13--17, 2019,
San Francisco, CA, USA}
\acmPrice{}
\acmDOI{10.1145/3308558.3313497}
\acmISBN{978-1-4503-6674-8/19/05}
\begin{document}
\title{Feature Generation by Convolutional Neural Network for Click-Through Rate Prediction}

\author{Bin Liu$^1$, Ruiming Tang$^\ddagger$ $^1$, Yingzhi Chen$^\star$ $^2$, Jinkai Yu$^1$, Huifeng Guo$^1$, Yuzhou Zhang$^1$}
\thanks{$^\ddagger$: Corresponding author}
\thanks{$^\star$: This work was done when Yingzhi Chen was an intern in Huawei.\vspace{-2ex}}
\affiliation{%
  \institution{$^1$Noah's Ark Lab, Huawei, Shenzhen, China\\$^2$Jinan University, Guangzhou, China\\
  $^1$\{liubin131, tangruiming, yujinkai, huifeng.guo, zhangyuzhou3\}@huawei.com\\
  $^2$chen\_yingzhi@163.com}
}

\begin{spacing}{1.0}
\begin{abstract}
Click-Through Rate prediction is an important task in recommender systems, which aims to estimate the probability of a user to click on a given item. 
Recently, many deep models have been proposed to learn low-order and high-order feature interactions from original features. However, since useful interactions are always sparse, it is difficult for DNN 
to learn them effectively under a large number of parameters. In real scenarios, artificial features are able to improve the performance of deep models (such as Wide \& Deep Learning), but feature engineering is expensive and requires domain knowledge, making it impractical in different scenarios. 
Therefore, it is necessary to augment feature space automatically.
In this paper, We propose a novel Feature Generation by Convolutional Neural Network (FGCNN) model  
with two components: \textit{Feature Generation} and \textit{Deep Classifier}. \textit{Feature Generation} leverages the strength of CNN to generate local patterns and recombine them to generate new features. \textit{Deep Classifier} adopts the structure of IPNN to learn interactions  from the augmented feature space.   Experimental results on three large-scale datasets show that FGCNN significantly outperforms nine state-of-the-art models. Moreover, when applying some state-of-the-art models as \textit{Deep Classifier}, better performance is always achieved, showing the great compatibility of our FGCNN model. This work explores a novel direction for CTR predictions: it is quite useful to reduce the learning difficulties of DNN by automatically identifying important features.
\end{abstract}

%


\maketitle

\input{introduction_1023_huifeng}

\begin{figure*}[t]
\vspace{-1ex}
\includegraphics[width=0.85\textwidth]{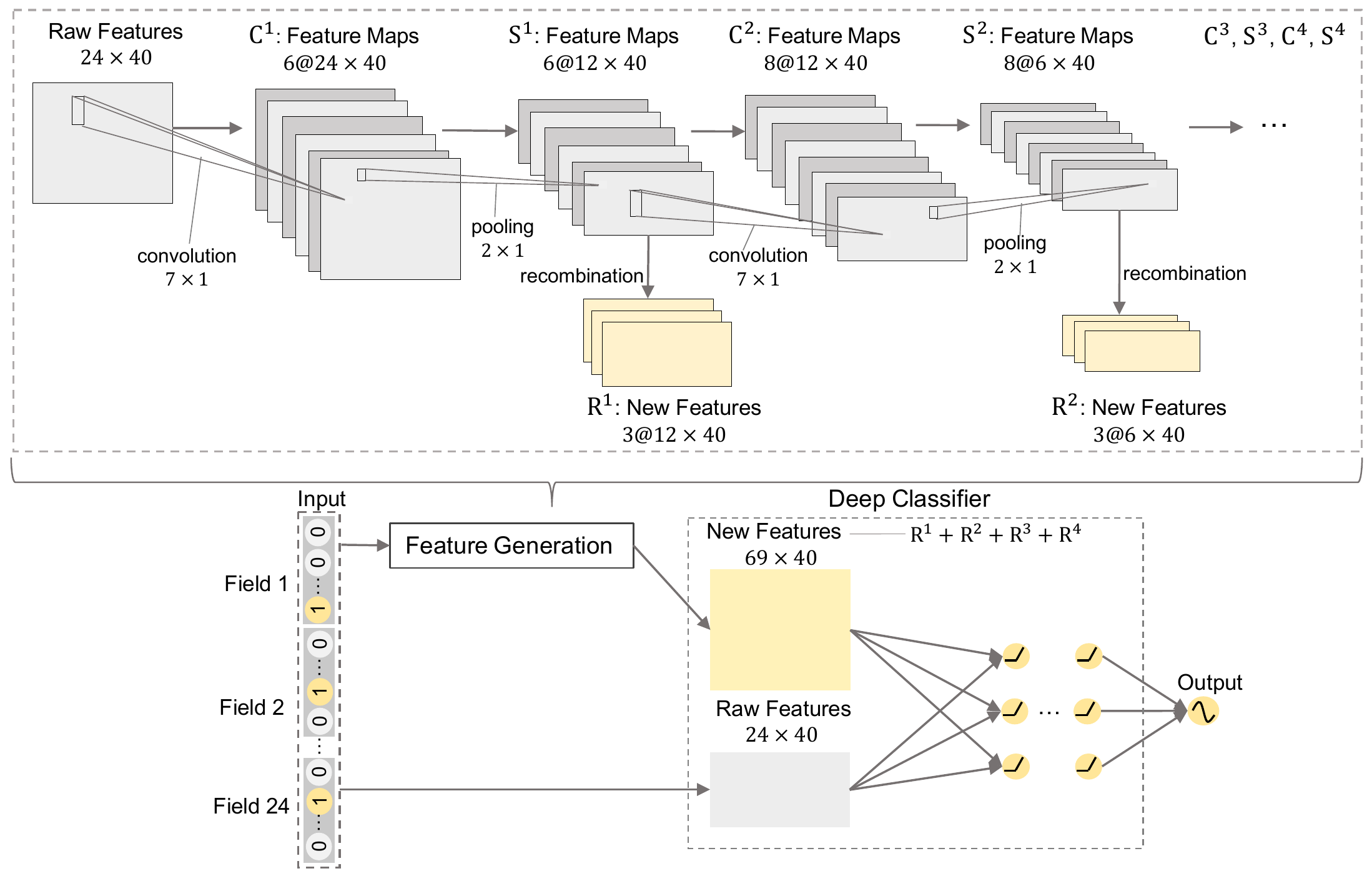}
\vspace{-2ex}
\caption{An overview of Feature Generation by Convolutional Neural Network Model
 (The hyper-parameters in the figure are the best setting of FGCNN on Avazu Dataset)}
 \vspace{-2ex}
\label{fig:model_overview}
\end{figure*}

\input{model_1026_ruiming}

\input{experiments_1026_huifeng}

\input{related_work}

\section{Conclusion}

 In this paper, we propose a FGCNN model for CTR prediction which aims to reduce the learning difficulties of DNN models by identifying important features in advance. The model consists of two components: \textit{Feature Generation} and \textit{Deep Classifier}. \textit{Feature Generation} leverages the strength of CNN to identify useful local patterns and it alleviates the weakness of CNN by introducing a \textit{Recombination Layer} to generate global new features from the recombination of the local patterns. In \textit{Deep Classifier}, most existing deep models can be applied on the augmented feature space. 
 Extensive experiments are conducted on three large-scale datasets where the results show that FGCNN outperforms nine state-of-the-art models. Moreover, when applying other models in \textit{Deep Classifier}, compared with the original model without \textit{Feature Generation}, better performance is always achieved, which demonstrates the effectiveness of the generated features. Step-by-step experiments show that each component in FGCNN contributes to the final performance. Furthermore, compared with the traditional CNN structure, our CNN+Recombination structure in \textit{Feature Generation} always performs better and more stable when shuffling the arrangement order of raw features. This work explores a novel direction for CTR prediction that it is  effective to automatically generate important features first rather than feeding the raw embeddings to deep learning models directly.
 \newpage
\end{spacing}
 \bibliographystyle{ACM-Reference-Format}
 \balance
\bibliography{main}

\end{document}

%% file: introduction_1023_huifeng.tex
\section{Introduction}
\label{Introduction}

Click-Through Rate (CTR) is a crucial task for recommender systems, which estimates the probability of a user to click on  a given item~\cite{pin,deepfm}. In an online advertising application, which is a billion-dollar scenario, the ranking strategy of candidate advertisements is by CTR$\times$bid where ``bid'' is the profit that the system receives once the advertisement is clicked on. In such applications, the performance of CTR prediction models~\cite{pin,deepfm,din} is one of the core factors determining system's revenue.
\begin{figure}[!t]
 \includegraphics[width=0.48\textwidth]{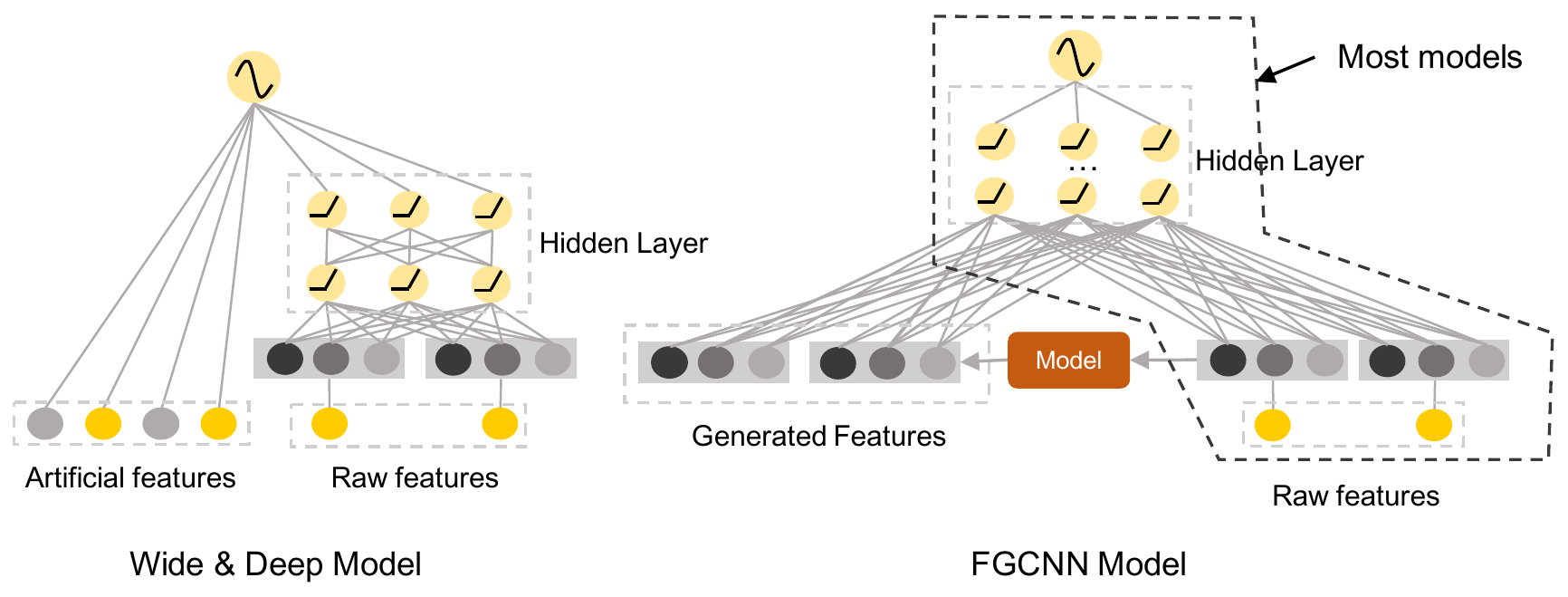}
 \vspace{-6ex}
 \caption{Comparision Between our Model and Wide \& Deep Model}\label{fig:wide-deep}
 \vspace{-3ex}
\end{figure}

The key challenge for CTR prediction tasks is to effectively model feature interactions. Generalized linear models, such as FTRL~\cite{ftrl}, perform well in practice, but these models lack the ability to learn feature interactions.
To overcome the limitation, Factorization Machine~\cite{fm} and its variants~\cite{ffm} are proposed to model pairwise feature interactions as the inner product of latent vectors and show promising results.
Recently, deep neural networks (DNN) have achieved remarkable progress in computer vision~
\cite{Kaiming2015Deep,Szegedy2016Inception} 
and natural language processing
~\cite{Bahdanau2014Neural,Vaswani2017Attention}. And some deep learning models have been proposed for CTR predictions, such as PIN~\cite{pin}, xDeepFM~\cite{xdeepfm} and etc.
Such models feed raw features to a deep neural network to learn feature interactions explicitly or implicitly. Theoretically, DNN is able to learn arbitrary feature interactions from the raw features. However, due to that useful interactions are ususally sparse compared with the combination space of raw features, it is of great difficulties to learn them effectively from a large number of parameters~\cite{pin,Shalevshwartz2017Failures}.

Observing such difficulties, Wide \& Deep Learning~\cite{cheng2016wide} leverages feature engineering in the \emph{wide} component to help the learning of \emph{deep} component.  With the help of artificial features, the performance of deep component is improved significantly (0.6\% improvement on offline AUC and 1\% improvement on online CTR).
However, feature engineering can be expensive and requires domain knowledge. If we can generate sophisticated feature interactions automatically by machine learning models, it will be more practical and robust.

Therefore, as shown in Figure~\ref{fig:wide-deep}, we propose a general framework for automatical feature generation. 
Raw features are input into a machine learning model (represented by the red box in Figure~\ref{fig:wide-deep}) to identify and generate new features \footnote{Here, new features are the feature interactions of the raw features. In the rest of this paper, we may use the term ``new features'' for the ease of presentation.}. After that, the  raw features and the generated new features are combined and fed into a deep neural network.  The generated features are able to reduce the learning difficulties of deep models by capturing the sparse but important feature interactions in advance.

The most straightforward way for automatical feature generation is to perform Multi-Layer Perceptron (MLP\footnote{MLP is a neural network with several fully connected layers.}) and use the hidden neurons as the generated features. However, as mentioned above, due to that useful feature interactions are usually sparse~\cite{pin}, it is rather difficult for MLP to learn such interactions from a huge parameter space. For example, suppose we have four user features: \textit{Name}, \textit{Age}, \textit{Height}, \textit{Gender} to predict whether a user will download an online game. Assume that  the feature interaction between \textit{Age} and \textit{Gender} is the only signal that matters, so that an optimal model should identify this and only this feature interaction. When performing MLP with only one hidden layer, the optimal weights associated to the embeddings of \textit{Name} and \textit{Height} should be all 0's, which is fairly difficult to achieve. 

As an advanced neural network structure, Convolutional Neural Network (CNN) 
has achieved great success in the area of computer vision~\cite{Huang2016Densely} and natural language processing~\cite{Bahdanau2014Neural}. In CNN, the design of shared weights and pooling mechanism greatly reduces the number of parameters needed to find important local patterns and it will alleviate the optimization difficulties of  later MLP structures. Therefore, CNN provides a potentially good solution to realize our idea (identify the sparse but important feature interactions). However, applying CNN  directly could result in unsatisfactory performance. In CTR prediction, different arrange orders of original features do not have different meanings. For example, whether the arrangement order of features being \emph{(Name, Age, Height, Gender)} or \emph{(Age, Name, Height, Gender)} does not make any difference to describe the semantics of a sample, which is completely different from the case of images and sentences. If we only use the neighbor patterns extracted by CNN, many useful global feature interactions will be lost.
This is also why CNN models do not perform well for CTR prediction task. 
To overcome this limitation, we perform CNN and MLP, which complement each other, to learn global-local feature interactions for feature generation.

In this paper, we propose a new model for CTR prediction task, namely Feature Generation by Convolutional Neural Network (FGCNN), which consists of two components: \emph{Feature Generation} and \emph{Deep Classifier}. 
In \emph{Feature Generation}, a CNN+MLP structure is designed to identify and generate new important features from raw features. More specifically, CNN is performed to learn neighbor feature interactions, while MLP is applied to recombine them  to extract global feature interactions. After \emph{Feature Generation}, the feature space can be augmented by combining raw features and new features.
In \emph{Deep Classifier}, almost all state-of-the-art network structures (such as PIN~\cite{pin}, xDeepFM~\cite{xdeepfm}, DeepFM~\cite{deepfm}) can be adopted. Therefore, our model has good compatibility with the state-of-the-art models in recommender systems. 
For the ease of illustration, we will adopt IPNN model~\cite{pin,pnn} as \emph{Deep Classifier} in FGCNN, due to its good trade-off between model complexity and accuracy.  Experimental results in three large-scale datasets show that FGCNN significantly outperforms nine state-of-the-art models, demonstrating the effectiveness of FGCNN. While adopting other models in \emph{Deep Classifier}, better performance is always achieved, which shows the usefulness of the generated features. Step-by-step analyses show that each component in FGCNN contributes to the final performance. Compared with traditional CNN structure, our CNN+MLP structure performs better and more stable when the order of raw features changes, which demonstrates the robustness of FGCNN.

To summarize, the main contributions of this paper can be highlighted as follows:
\begin{itemize}
\vspace{-1ex}
\item   An important direction is identified for CTR prediction: it is both necessary and useful to reduce the optimization difficulties of deep learning models by automatically generating important features in advance.

\item We propose a new model-FGCNN for automatical feature generation and classification, which consists of two components: \emph{Feature Generation} and \emph{Deep Classifier}. \emph{Feature Generation} leverages CNN and MLP, which complement each other, to  identify the important but sparse features. Moreover, almost all other CTR models can be applied in \emph{Deep Classifier} to learn and predict based on the generated and raw features.

\item Experiments on three large-scale datasets demonstrate the overall effectiveness of FGCNN model. When the generated features are used in other models, better performance is always achieved, which shows the great compatibility and robustness of our FGCNN model.
\end{itemize}
The rest of this paper is organized as follows: Section 2 presents the proposed FGCNN model in detail. Experimental results will be shown and discussed in Section 3. Related works will be introduced in Section 4. Section 5 concludes the paper. 

%% file: model_1026_ruiming.tex
\section{Feature Generation by Convolutional Neural Network Model}

\subsection{Overview}

In this section, we will describe the proposed Feature Generation by Convolutional Neural Network (FGCNN) model in detail. The used notations are summarized in Table~\ref{table:notations}.

As shown in Figure~\ref{fig:model_overview}, FGCNN model consists of two components: \textit{Feature Generation} and \textit{Deep Classifier}. More specifically, \textit{Feature Generation} focuses on identifying useful local and global patterns to generate new features as a complement to  raw features, while \textit{Deep Classifier} learns and predicts based on the augmented feature space  through a deep learning model. Besides the two components, we will also formalize \textit{Feature Embedding} of our model. The details of these components are presented in the following subsections. 

\subsection{Feature Embedding}\label{embedding}

In most CTR prediction tasks, data is collected in a multi-field categorical form\footnote{Features in numerical form are usually transformed into categorical form by bucketing.}~\cite{Shan2016Deep,Zhang2016Deep}, so that each data instance is normally transformed into a high-dimensional sparse (binary) vector via one-hot encoding~\cite{He2014Practical}. For example,
(Gender=Male, Height=175, Age=18, Name=Bob) can be represented as:
 \begin{equation}
  \underbrace{(0,1)}_{\text{Gender=Male}}\underbrace{ (0,...,1,0,0)}_{\text{Height=175}}\underbrace{(0,1,...,0,0)}_{\text{Age=18}}\underbrace{(1,0,0,..,0)}_{\text{Name=Bob}}
 \end{equation}
 
An embedding layer is applied upon the raw features' input to compress them to low-dimensional vectors. In our model, if a field is univalent (e.g., ``Gender=Male''), its embedding is the feature embedding of the field; if a field is multivalent (e.g., ``Interest=Football, Basketball''), the embedding of the field takes the sum of features' embeddings~\cite{dnnyoutube}.

More formally, in an instance, each field $i\ (1 \le i \le n_f)$ is represented as a low-dimensional vector $e_i \in R^{1\times k}$, where $n_f$ is the number of fields and $k$ is embedding size.
Therefore, each instance can be represented as an embedding matrix $E = (e_1^T, e_2^T, ..., e_{n_f}^T)^T$, where $E \in R^{n_f\times k}$. 
In FGCNN model, the embedding matrix $E$ can be utilized in both \textit{Feature Generation} and \textit{Deep Classifier}.  To avoid the inconsistency of gradient direction when updating parameters, we will introduce another embedding matrix $E' \in R^{n_f\times k}$ for \emph{Deep Classifier} while $E$ is used in \emph{Feature Generation}.
\begin{table}[t]
\vspace{-1ex}
\caption{Notations}
\vspace{-3ex}
\label{table:notations}
\resizebox{0.45\textwidth}{!}{
\begin{tabular}{|c|l|}
\hline
Parameter & Meaning \\ \hline
$k$ & embedding size \\ \hline
$t_f$ & the total number of one-hot features \\ \hline
$n_f$ & the number of fields\\ \hline
$n_c$ & the number of convolutional layers \\ \hline
$n_h$ & the number of hidden layers \\ \hline
$h^i$ & the height of convolutional kernel in the $i$-th convolutional layer \\ \hline
$h_p$ & pooling height (width=1) of pooling layers \\ \hline
$e_i$ &  the embedding vector for $i$-th field \\ \hline
$E^i$ & the input of the $i$-th convolutional layer \\ \hline
$m_c^i$ & the number of feature maps in the $i$-th convolutional layer \\ \hline
$m_r^i$ & the number of new features' map in the $i$-th recombination layer \\ \hline
$N_i$ & the number of generated features in the $i$-th recombination layer (=$n_f/h_p^im_r^i$) \\ \hline
$C_{:,:,j}^{i}$ & the $j$-th output  feature map of the $i$-{th} convolutional layer \\ \hline
$\mathbb{WC}_{:,:,j}^{i}$ & weights for the $j$-th output feature map of $i$-th convolutional layer\\ \hline
$S_{:,:,j}^{i}$ & the $j$-th output feature map of the $i$-th pooling layer \\ \hline
$\mathbb{WR}^{i}$ & weights for the $i$-th recombination layer \\ \hline
$\mathbb{BR}^{i}$ & bias for the $i$-th recombination layer \\ \hline
$R_{:,:,j}^{i}$ & the $j$-th output map of new features in the $i$-th recombination layer \\ \hline
$W^{i}$ & weights for the $i$-th hidden layer\\ \hline
$B^{i}$ & biases for the $i$-th hidden layer \\ \hline
$H_i$ & number of hidden neorons in the $i$-th hidden layer \\ \hline
\end{tabular}}
\vspace{-3ex}
\end{table}
\subsection{Feature Generation}

As stated in Section~\ref{Introduction}, generating new features from raw features helps improve the performance of deep learning models (as demonstrated by Wide \& Deep Learning~\cite{cheng2016wide}). To achieve this goal, \emph{Feature Generation} designs a proper neural network structure to identify useful feature interactions then generate new features automatically. As argued in Section~\ref{Introduction}, using MLP or CNN alone is not able to generate effective feature interactions from raw features, due to the following reasons: \emph{Firstly}, 
 useful feature interactions are always sparse in the combination space of raw features. Therefore, it is difficult for MLP to learn them from a large amount of parameters. \emph{Secondly}, although CNN can alleviates optimization difficulties of MLP by reducing the number of paramters, it only generates neighbor feature interactions which can lose many useful global feature interactions.

 \begin{figure}[t]
  \includegraphics[width=0.45\textwidth]{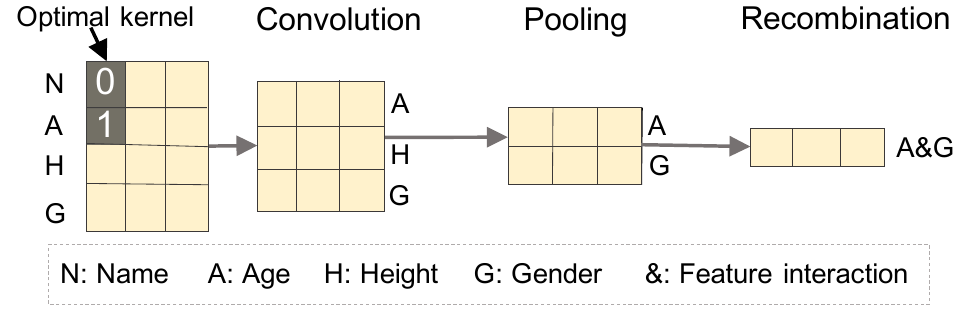}
  \vspace{-3ex}
  \caption{CNN+Recombination structure is able to capture global non-neighbor feature interactions to generate new features. CNN consists of convolutional layer and pooling layer, while Recombination consists of a fully connected layer.}
  \vspace{-5ex}
  \label{fig:CNN+Pooling+MLP}
 \end{figure}

In order to overcome the weakness of applying MLP or CNN alone, as shown in the upper part of Figure~\ref{fig:model_overview}, we perform CNN and MLP~\footnote{According to its function, we will call MLP with one hidden layer as recombination layer later.} as a complement to each other for feature generation. Figure~\ref{fig:CNN+Pooling+MLP} shows an example of CNN+Recombination structure to capture global feature interactions. As can be observed, CNN learns useful neighbor feature patterns with a limited number of parameters, while recombination layer (which is a fully connected layer) generates global feature interactions based on the neighbor patterns provided by CNN. Therefore, important features can be generated  effectively via this neural network structure, which has fewer parameters than directly applying MLP for feature generation. 

In the following parts, we detail the CNN+Recombination structure of \textit{Feature Generation}, namely, \textit{Convolutional Layer}, \textit{Pooling layer}  and \textit{Recombination layer}.

\subsubsection{Convolutional Layer}
 
Each instance is represented as an embedding matrix $E\in R^{n_f\times k}$ via feature embedding, where $n_f$ is the number of fields and $k$ is embedding size. For convenience, reshape the embedding matrix as $E^{1} \in R^{n_f\times k\times 1}$ as the input matrix of the first convolutional layer.
To capture the neighbor feature interactions, a convolutional layer is obtained by convolving a matrix $\mathbb{WC}^1 \in R^{h^1\times 1\times 1\times m_c^{1}}$ with non-linear activation functions (where $h^1$ is the height of the first convolutional weight matrix and $m_c^1$ is the number of feature maps in the first convolutional layer). Suppose the output of the first convolutional layer is denoted as $C^{1} \in R^{n_f\times k\times m_c^1}$, we can formulate the convolutional layer as follows:
\begin{shrinkeq}{0ex}
\begin{equation}
 C_{p,q,i}^1 = tanh(\sum_{m=1}^1\sum_{j=1}^{h^1} E_{p+j-1,q,m}^1 {\mathbb{WC}}_{j,1,1,i}^1)
\end{equation}
\begin{equation}
 tanh(x) = \frac{exp(x)-exp(-x)}{exp(x)+exp(-x)}
\end{equation}

\end{shrinkeq}

where $C_{:,:,i}^1$ denotes the $i$-th feature map in the first convolutional layer and $p, q$ are the row and column index of the $i$-th feature map. Notice that the above equation excludes padding which is performed in practice.

\subsubsection{Pooling Layer}

After the first convolutional layer, a max-pooling layer is applied to seize the most important feature interactions and reduce the number of parameters. We refer $h_p$ as the height of pooling layers (width=1). 
The output in the first pooling layer is $S^1 \in R^{(n_f/h_p)\times k\times m_c^1}$: 
\begin{equation}
  S^1_{p,q,i} = max(C^1_{p\cdot h_p,q,i},\ ...\ , C^1_{p\cdot h_p+h_p-1,q,i})
\end{equation}
The pooling result of the $i$-th pooling layer will be the input for the $(i+1)$-th convolutional layer:
$ E^{i+1} = S^i$.

\subsubsection{Recombination Layer}

After the first convolutional layer and pooling layer, $S^1 \in R^{(n_f/h_p)\times k\times m_c^1}$ contains the patterns of  neighbor features. Due to the nature of CNN, global non-neighbor feature interactions will be ignored if $S^1$ is regarded as the generated new features. Therefore, we introduce a fully connected layer to recombine the local neighbor feature patterns and generate important new features. We denote the weight matrix as $\mathbb{WR}^1\in R^{(n_f/h_pkm_c^1) \times  (n_f/h_pkm_r^1)}$ and the bias as ${\mathbb{BR}^1} \in R^{(n_f/h_pkm_r^1)}$, where $m_c^1$ is the number of feature maps in  the first convolutional layer and $m_r^1$  is the number of new features' map in the first recombination Layer. Therefore, in the $i$-th recombination layer, $n_f/h_p^im_r^i$ features are generated:
\begin{shrinkeq}{-1ex}
\begin{align}
  &R^1 = tanh(S^1\cdot \mathbb{WR}^1 + {\mathbb{BR}^1})
\end{align}
\end{shrinkeq}
 
\subsubsection{Concatenation}
\vspace{-0.5ex}
New features can be generated by performing CNN+Recombination multiple times. Assume there are $n_c$ convolutional layers, pooling layers and recombination layers and $N_i =n_f/h_p^im_r^i$ fields of  features are generated by $i$-th round denoted as $R^i$. The overall new features $\mathcal{R} \in R^{N\times k}$ (where $N=\sum_{i=1}^{n_c}{N_i}$) generated by \textit{Feature Generation} are formalized as:
\vspace{-0.5ex}
\begin{shrinkeq}{-1ex}
 \begin{align}
\mathcal{R}&= (R^1, R^2, ..., R^{n_c})
 \end{align}
 \end{shrinkeq}

Then, raw features and new features are concatenated as
\begin{shrinkeq}{-1ex}
\begin{align}
\mathbb{E}& = (E'^T,\mathcal{R}^T)^T
\end{align}
\end{shrinkeq}

where $E'$ is the embedding matrix of raw features  for \textit{Deep Classifier} (see Section~\ref{embedding}). Both the raw features and new features are utilized for CTR prediction in \textit{Deep Classifier}, which will be elaborated in the next subsection.

\subsection{Deep Classifier}\label{sec:clf}

As mentioned above,  raw features and new features are concatenated as an augmented embedding matrix $\mathbb{E} \in R^{(N+n_f)\times k}$, where $N$ and $n_f$ are the number of fields of new features and raw features respectively. $\mathbb{E}$ is input into \emph{Deep Classifier}, which aims to learn further interactions between the raw features and new generated features. In this subsection, for the ease of presentation, we adopt IPNN model~\cite{pin} as the network structure in \textit{Deep Classifier}, due to its good trade-off between model complexity and accuracy. In fact, any advanced network structure can be adopted, which shows the compatibility of FGCNN with the existing works. The compatibility of FGCNN model will be verified empirically in  Section 3. 

\subsubsection{Network Structure}

IPNN model~\citep{pin} combines the learning ability of FM~\cite{Rendle2011Factorization} and MLP. It utilizes an FM layer to extract  pairwise feature interactions from embedding vectors by inner product operations. After that,   the embeddings of input features and the results of FM layer are concatenated and fed to MLP for learning. As evaluated in~\cite{pin}, the performance of IPNN is slightly worse than PIN (the best model in~\cite{pin}), but IPNN is much more efficient.  We will illustrate the network structure of IPNN model.
\begin{figure}[t]
 \centering
 \includegraphics[width=0.48\textwidth]{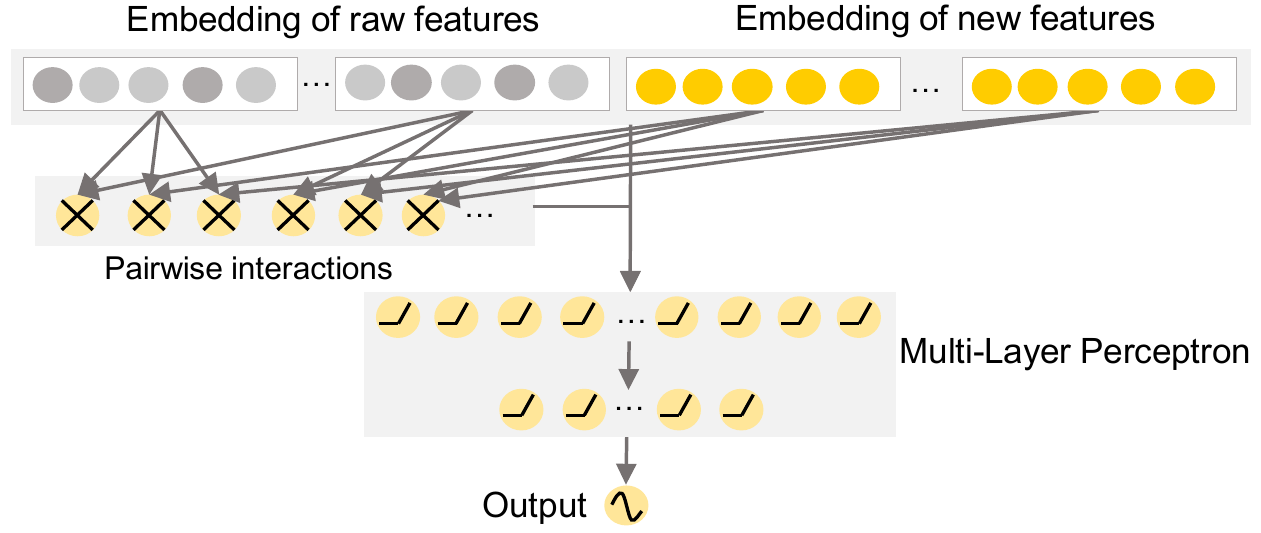}
 \vspace{-5ex}
 \caption{Structure of IPNN Model}
 \vspace{-4ex}
  \label{fig:ipnn}
\end{figure}

As shown in Figure~\ref{fig:ipnn}, the pariwise feature interactions of augmented embedding matrix $\mathbb{E}\in R^{(N+n_f)\times k}$ are modeled by an FM layer, as follows:
\vspace{-0.5ex}
\begin{equation}
    R_{fm} = (<\mathbb{E}_{1},\mathbb{E}_2>,...,<\mathbb{E}_{N+n_f-1},\mathbb{E}_{N+n_f}>)
\end{equation}
where $\mathbb{E}_{i}$ is the embedding of the $i$-th field, $<a,b>$ means the inner product of $a$ and $b$. The number of pairwise feature interactions in the FM layer is $\frac{(N+n_f)(N+n_f-1)}{2}$.

After the FM layer, $R_{fm}$ is concatenated with the augmented embedding matrix $\mathbb{E}$, which are fed into MLP with $n_h$ hidden layers to learn implicit feature interactions. We refer the input for the $i$-th hidden layer as $I_i$ and the output for the $i$-th hidden layer as $O_i$. The MLP is formulated as:
\begin{shrinkeq}{-0.1ex}
\begin{align}
 &I_1 = (R_{fm}, flatten(\mathbb{E}))\\
 &O_i = relu(I_{i}W^i+B^i) \\
 &I_{i+1} = O_{i}
\end{align}
\end{shrinkeq}
where $W^i$ and $B^i$ are the weight matrix and bias of the $i$-th hidden layer in MLP. For the last hidden layer (the $n_h$-th layer), we will make final predictions:
\begin{equation}
 \hat{y} = sigmoid(O_{n_h}W^{n_h+1}+b^{n_h+1})
\end{equation}
\vspace{-5ex}
\subsubsection{Batch Normalization}

Batch Normalization (BN) is proposed in~\cite{BN} to solve covariant shift and accelerate DNN training. It normalizes activations $w^Tx$ using statistics within a mini-batch, as follows:
\begin{equation}
 BN(w^Tx) = \frac{w^Tx-avg_i(w^Tx)}{std_i(w^Tx)} g +b
\end{equation} 
where $g$ and $b$ scale and shift the normalized values.

In the FGCNN model, Batch Normalization is applied before each activation function to accelerate model training.

\subsubsection{Objective Function}

The objective function in the FGCNN model is to minimize cross entropy of predicted values and the labels, which is defined as:
\begin{equation}
 \mathcal{L}(y,\hat{y}) = -y log\hat{y}-(1-y)log(1-\hat{y})
\end{equation}
where $y \in \{0,1\}$ is the label and $\hat{y} \in (0,1)$ is the predicted probability of $y=1$.

\subsection{Complexity Analysis}
\label{complexity}
In this section, we analyse the space and time complexity of FGCNN model. Notations in Table ~\ref{table:notations} will be reused.  

\subsubsection{Space Complexity}

The space complexity of FGCNN model consists of  three parts, namely \textit{Feature Embedding}, \textit{Feature Generation} and \textit{Deep Classifier}. In feature embedding, since there are two matrices (one for \textit{Feature Generation} and one for \textit{Deep Classifier} (as discussed in  Section~\ref{embedding})), there are $s_0 = 2t_fk$ parameters, where $t_f$ is the total number of one-hot raw features.

In \textit{Feature Generation}, there are $h^im_c^{i-1}m_c^i$ parameters for convolutional matrix in $i$-th convolutional layer and $n_f/h_p^ik m_c^i$ neurons in the output of $i$-th pooling layer. In addition, the number of parameters in $i$-th recombination layer is $n_f^2/h_p^{2i}k^2m_c^im_r^i$, which will generate $N_i = n_f/h_p^im_r^i$ fields of new features ($n_f/h_p^i = N_i/m_r^i$). Therefore, the total parameters of \textit{Feature Generation} is:
\begin{shrinkeq}{-1ex}
\begin{align}
&\sum_{i=1}^{n_c} h^im_c^{i-1}m_c^i+ n_f^2/h_p^{2i}k^2m_c^im_r^i = \\
&\sum_{i=1}^{n_c} h^im_c^{i-1}m_c^i + N_i^2k^2m_c^i/m_r^i
\end{align}
\end{shrinkeq}
After the \textit{Feature Generation} process, we can get $T = n_f+\sum_{i=1}^{n_c} N_i$ feature embeddings in total.
Normally, $m_r^i$ is usually set as $1,2,3,4$ and  $m_r^i=m_r^{i-1}$. Meanwhile, $h_p$ is usually set to 2 so that $N_{i} = N_1/2^{i-1}$. Furthermore, $N_i^2 k^2 >> h_i m_c^{i-1}$ and $\sum_{i=1}^{n_c} m_c^i/m_r^i 2^{2(i-1)}$ is usually a small number. The space complexity $s_1$ of the feature generating process can be simplified as:
\begin{shrinkeq}{-0.5ex}
\begin{align}
s_1&=O(\sum_{i=1}^{n_c} N_1^2/2^{2(i-1)}k^2m_c^i/m_r^i) = O(N_1^2 k^2)
\end{align}
\end{shrinkeq}

For the first hidden layer of \textit{Deep Classifier}, the number of parameters in the weight matrix is $(T(T-1)/2+Tk)H_1$ (recall that the input is the raw and new features' embedding and their pairwised product). In the $i$-th hidden layer, the number of parameters is $H_{i-1}H_i$. Therefore, the space complexity of \textit{Deep Classifier} is:
\begin{shrinkeq}{-0.5ex}
\begin{equation}\label{eq:p-dpl}
s_2=O(\frac{T(T-1)+2Tk}{2}H_1 + \sum_{i=2}^{n_h} H_iH_{i-1})
\end{equation}
\end{shrinkeq}
where Eq.~(\ref{eq:p-dpl}) can be simplified as  $O(T^2H_1+\sum_{i=2}^{n_h}H_iH_{i-1})$  since $Tk < T(T-1)$ usually. 

In summary, the overall space complexity of FGCNN is $s_0+s_1+s_2=O(t_fk+N_1^2k^2+T^2H_1+\sum_{i=1}^qH_iH_{i-1})$, which is dominated by the number of one-hot features, the number of generated features, the number of hidden neurons and the embedding size.

\subsubsection{Time Complexity}

We first analyze the time complexity in \textit{Feature Generation}. In the $i$-th convolutional layer, the output dimension is $n_f/h_p^{i-1}km_c^i = N_ih_p/m_r^ikm_c^i$ (recall that $N_i = n_f/h_p^im_r^i$) and each point is calculated from $h^im_c^{i-1}$ data points where $h^i$ is the  height of convolutional kernels in the $i$-th convolutional layer. Therefore, the time complexity of the $i$-th convolutional layer is $N_i h_p km_c^i  h^i m_c^{i-1}/m_r^i$ . In the pooling layer, the time complexity of the $i$-th pooling layer is $n_f/h_p^{i-1}km_c^i=N_ih_pkm_c^i/m_r^i $. In the $i$-th recombination layer, the time complexity is $n/h_p^ikm_c^i \cdot  N_ik = N_i^2k^2m_c^i/m_r^i$ (recall $h_p =2$ and $N_i = N_1 / 2^{i-1}$). Therefore, total time complexity of the \textit{Feature Generation} is
\begin{shrinkeq}{-1ex}
\begin{align*}
 t_1 &= \sum_{i=1}^{n_c} N_i h_p km_c^i  h^i m_c^{i-1}/m_r^i+ N_ih_pkm_c^i/m_r^i + N_i^2k^2m_c^i/m_r^i\\ 
 &= \sum_{i=1}^{n_c} N_1km_c^{i}/(m_r^i2^{i-1})(2 h^i m_c^{i-1}+2 + N_1/2^{i-1}k)\\
 &= O(N_1k\sum_{i=1}^{n_c}m_c^{i-1}h^i+N_1^2k^2)\ 
\end{align*}
\end{shrinkeq}
(Notice: $m_c^{i}/(m_r^i2^{i})$ is usually a small number ranging from 2 to 10)

Then, we analyze the time complexity of \textit{Deep Classifier} (taking IPNN as an example). In the FM layer, the time complexity to calculate pairwise inner product is $\frac{T(T-1)k}{2}$. The number of neurons, that are input to the first hidden layer is $T(T-1)/2+Tk$. Therefore, the time complexity of the first hidden layer is $O((T(T-1)/2+Tk)H_1)=O(T^2H_1)$. For the other hidden layers, the time complexity is $O(H_iH_{i-1})$. Therefore, the total time complexity in \textit{Deep Classifier} is
\begin{shrinkeq}{-1ex}
\begin{equation}
 t_2 = O(T^2H_1)+\sum_{i=2}^{n_h} O(H_iH_{i-1})
\end{equation}
\end{shrinkeq}

In summary, the overall time complexity for FGCNN model is $t_1+t_2 = O(N_1k\sum_{i=1}^{n_c}m_c^{i-1}h^i+N_1^2k^2+T^2H_1+\sum_{i=2}^qH_iH_{i-1})$ which is dominated by the number of generated features, the number of neurons in the hidden layers, embedding size and convolutional parameters.

%% file: experiments_1026_huifeng.tex
\section{Experiments}

In this section, we conduct extensive experiments to answer the following questions.
\begin{itemize}
       \item \textbf{(Q1)} How does FGCNN perform, compared to the state-of-the-art models for CTR prediction task?
       \item \textbf{(Q2)} Can FGCNN improve the performance of other state-of-the-art models by using their structures in  \emph{Deep Classifier}?
       \item \textbf{(Q3)} How does each key structure in FGCNN boost the performance of FGCNN?
       \item \textbf{(Q4)} How do the key hyper-parameters of FGCNN
(i.e., size of convolutional kernels,number of convolutional layers, number
of generated features) impact its performance?
       \item \textbf{(Q5)} How does  \emph{Feature Generation}  perform when the order of raw features are randomly shuffled?
\end{itemize}

Note that when studying questions other than \textbf{Q2}, we adpot IPNN as the \emph{Deep Classifier} in FGCNN.

\subsection{Experimental Setup}
\begin{table}[t]
\caption{Dataset Statistics}
\vspace{-3ex}
\label{table:dataset}
\resizebox{0.45\textwidth}{!}{
\begin{tabular}{l|llll}
\hline
Dataset & \#instances & \#features & \#fields & pos ratio \\ \hline
Criteo & $1\times 10^{8}$ & $1\times 10^{6}$ & 39 & 0.5 \\
Avazu & $4\times 10^{7}$ & $6\times 10^{5}$ & 24 & 0.17 \\
Huawei App Store & $2.3\times 10^{8}$ & $1.6\times 10^{5}$ & 29 & 0.05 \\\hline
\end{tabular}
}
\vspace{-1ex}
\end{table}
\begin{table}[h]
\vspace{-3ex}
\caption{Parameter Settings}
\vspace{-3ex}
\label{table:parameter}
\scriptsize
\resizebox{0.45\textwidth}{!}{
\begin{tabular}{|l|l|l|l|}
\hline
Param   & Criteo                                                                                                                       & Avazu                                                                                                                               & Huawei App Store                                                                                                                                            \\ \hline
General & \begin{tabular}[c]{@{}l@{}}bs=2000\\ opt=Adam\\ lr=1e-3\end{tabular}                                                         & \begin{tabular}[c]{@{}l@{}}bs=2000\\ opt=Adam\\ lr=1e-3\end{tabular}                                                                & \begin{tabular}[c]{@{}l@{}}bs=1000\\ opt=Adam\\ lr=1e-4\\ l2=1e-6\end{tabular}                                                                     \\ \hline
LR      & --                                                                                                                           & --                                                                                                                                  & --                                                                                                                                                 \\ \hline
GBDT    & \begin{tabular}[c]{@{}l@{}}depth=25\\ \#tree=1300\end{tabular}                                                               & \begin{tabular}[c]{@{}l@{}}depth=18\\ \#tree=1000\end{tabular}                                                                      & \begin{tabular}[c]{@{}l@{}}depth=10\\ \#tree=1400\end{tabular}                                                                                     \\ \hline
FM      & k=20                                                                                                                         & k=40                                                                                                                                & k=40                                                                                                                                               \\ \hline
FFM     & k=4                                                                                                                          & k=4                                                                                                                                 & k=12                                                                                                                                               \\ \hline
CCPM    & \begin{tabular}[c]{@{}l@{}}k=20\\ conv= 7*1\\ kernel=[256]\\ net=[256*3,1]\end{tabular}                                      & \begin{tabular}[c]{@{}l@{}}k=40\\ conv: 7*1\\ kernel = [128]\\ net= [128*3,1]\end{tabular}                                          & \begin{tabular}[c]{@{}l@{}}k=40\\ conv= 13*1\\ kernel= [8,16,32,64]\\ net=[512,256,128,1]\\ drop= 0.8\end{tabular}                                  \\ \hline
DeepFM  & \begin{tabular}[c]{@{}l@{}}k=20\\ LN=T\\ net=[700*5,1]\end{tabular}                                                          & \begin{tabular}[c]{@{}l@{}}k=40\\ LN=T\\ net=[500*5,1]\end{tabular}                                                                 & \begin{tabular}[c]{@{}l@{}}k=40\\ net=[2048,1024,512,1]\\ drop=0.9\end{tabular}                                                                    \\ \hline
XdeepFM & \begin{tabular}[c]{@{}l@{}}k=20\\ net=[400*3,1]\\ CIN=[100*4]\end{tabular}                                                   & \begin{tabular}[c]{@{}l@{}}k=40\\ net=[700*5,1]\\ CIN:[100*2]\end{tabular}                                                          & \begin{tabular}[c]{@{}l@{}}k=40\\ net=[2048,1024,512,1]\\ drop=0.9\\ CIN:[100*4]\end{tabular}                                                                \\ \hline
IPNN    & \begin{tabular}[c]{@{}l@{}}k=20\\ LN=T\\ net=[700*5,1]\end{tabular}                                                          & \begin{tabular}[c]{@{}l@{}}k=40\\ LN=T\\ net=[500*5,1]\end{tabular}                                                                 & \begin{tabular}[c]{@{}l@{}}k=40\\ net=[2048,1024,\\ 512,256,128,1]\\ drop=0.7\end{tabular}                                                         \\ \hline
PIN     & \begin{tabular}[c]{@{}l@{}}k=20\\ LN=T\\ net=[700*5,1]\\ subnet=[40,5]\end{tabular}                                          & \begin{tabular}[c]{@{}l@{}}k=40\\ LN=T\\ net=[500*5,1]\\ sub-net=[40,5]\end{tabular}                                                & \begin{tabular}[c]{@{}l@{}}k=40\\ net=[2048,1024,512,1]\\ drop=0.9\\ sub-net=[80,10]\end{tabular}                                                  \\ \hline
FGCNN   & \begin{tabular}[c]{@{}l@{}}k=20\\ conv=9*1\\ kernel=[38,40,42,44]\\ new=[3,3,3,3]\\ BN=T\\ ruenet=[4096,2048,1]\end{tabular} & \begin{tabular}[c]{@{}l@{}}k=40\\ conv=7*1\\ kernel=[14,16,18,20]\\ new=[3,3,3,3]\\ BN=T\\ net=[4096,2048,\\ 1024,512,1]\end{tabular} & \begin{tabular}[c]{@{}l@{}}k=40\\ conv=13*1\\ kernel=[6, 8, 10, 12]\\ new=[2,2,2,2]\\ BN=T\\ net=[2048,1024,\\ 512,256,128,1]\\ drop=0.8\end{tabular} \\ \hline
\end{tabular}
}
Note: bs=batch size, opt=optimizer, lr=learning rate, l2=l2 regularization on Embedding Layer, k=embedding size, conv=shape of convolutional kernels, kernel=number of convolutional kernels, pool=max-pooling shape, net=MLP structure, sub-net=micro metwork, drop=dropout rate, LN=layer normalization, BN= batch normalization\{T=True\}, new=number of kernels for new features.
\vspace{-5ex}
\end{table}
\begin{table*}[t]
\centering
\vspace{-4ex}
\caption{Overall Performance\\
$\star: p < 10^{-2}\ \star\star: p<10^{-4} $ (two tailed t-test)}
\vspace{-3ex}
\label{table:overall}
\resizebox{0.8\textwidth}{!}{
\begin{tabular}{lll|ll|ll}
 \hline\hline
        & \multicolumn{2}{|c|}{Criteo}                                    & \multicolumn{2}{|c|}{Avazu}                                              & \multicolumn{2}{|c}{Huawei App Store}                        \\ \hline
\multicolumn{1}{l|}{Model}   & AUC                 & \multicolumn{1}{l|}{Log Loss}           & AUC(\%)                  & \multicolumn{1}{l|}{Log Loss}               & AUC                      & Log Loss                \\ \hline
\multicolumn{1}{l|}{LR}      & 78.00\%             & \multicolumn{1}{l|}{0.5631}             & 76.76\%                  & \multicolumn{1}{l|}{0.3868}                 & 90.12\%                  & 0.1371                  \\
\multicolumn{1}{l|}{GBDT}    & 78.62\%             & \multicolumn{1}{l|}{0.5560}             & 77.53\%                  & \multicolumn{1}{l|}{0.3824}                 & 92.68\%                  & 0.1227                  \\
\multicolumn{1}{l|}{FM}      & 79.09\%             & \multicolumn{1}{l|}{0.5500}             & 77.93\%                  & \multicolumn{1}{l|}{0.3805}                 & 93.26\%                  & 0.1191                  \\
\multicolumn{1}{l|}{FFM}     & 79.80\%             & \multicolumn{1}{l|}{0.5438}             & 78.31\%                  & \multicolumn{1}{l|}{0.3781}                 & 93.58\%                  & 0.1170                  \\
\multicolumn{1}{l|}{CCPM}    & 79.55\%             & \multicolumn{1}{l|}{0.5469}             & 78.12\%                  & \multicolumn{1}{l|}{0.3800}                 & 93.71\%                  & 0.1159                 \\
\multicolumn{1}{l|}{DeepFM}  & 79.91\%             & \multicolumn{1}{l|}{0.5423}             & 78.36\%                  & \multicolumn{1}{l|}{0.3777}                 & 93.91\%                  & 0.1145                  \\
\multicolumn{1}{l|}{xDeepFM} & 80.06\%             & \multicolumn{1}{l|}{0.5408}             & 78.55\%                  & \multicolumn{1}{l|}{0.3766}                 & 93.91\%                  & 0.1146                 \\
\multicolumn{1}{l|}{IPNN}    & 80.13\%             & \multicolumn{1}{l|}{0.5399}             & 78.68\%                  & \multicolumn{1}{l|}{0.3757}                 & \underline{93.95}\%                  & \underline{0.1143}                  \\
\multicolumn{1}{l|}{PIN}     & \underline{80.18\%}\footnotemark[11] & \multicolumn{1}{l|}{\underline{0.5393}} & \underline{78.72\%}      & \multicolumn{1}{l|}{\underline{0.3755}}     & 93.91\%      & 0.1146      \\ \hline
\multicolumn{1}{l|}{FGCNN}   & $\mathbf{80.22\%}^\star$     & \multicolumn{1}{l|}{$\mathbf{0.5388}^\star$}    & $\mathbf{78.83\%^{\star\star}}$ & \multicolumn{1}{l|}{$\mathbf{0.3746^{\star\star}}$} & $\mathbf{94.07\%^{\star\star}}$ & $\mathbf{0.1134^{\star\star}}$\\ \hline\hline
\end{tabular}
}
\vspace{-3ex}
\end{table*}

\subsubsection{Datasets} Experiments are conducted in the following three datasets:

\textit{Criteo}: Criteo\footnote{http://labs.criteo.com/downloads/download-terabyte-click-logs/} contains one month of click logs with billions of data samples. A small subset of Criteo was published in Criteo Display Advertising Challenge 2013 and FFM was the winning solution~\cite{pin, ffm}. We select ``day 6-12'' as training set while select ``day 13'' for evaluation. Due to the enormous data volume and serious class imbalance (i.e., only 3\% samples are positive), negative sampling is applied to keep the positive and negative ratio close to 1:1. We convert 13 numerical fields into one-hot features through bucketing, where the features in a certain field appearing less than 20 times are set as a dummy feature ``other``.

\textit{Avazu}: Avazu\footnote{http://www.kaggle.com/c/avazu-ctr-prediction} was published in the contest of Avazu Click-Through Rate Prediction, 2014. The public dataset is randomly splitted into training and test sets at 4:1. Meanwhile, we remove the features appearing less than 20 times to reduce dimensionality.

\textit{Huawei App Store}: In order to evaluate the performance in industrial CTR prediction problems, we conduct experiments on Huawei App Store Dataset. We collect users' click logs from Huawei App Store while logs from 20180617 to 20180623 are used for training and logs of 20180624 are used for test. Negative sampling is applied to reduce data amount and to adjust the ratio of positive class and negative class. The dataset contains app features (e.g., identification, category), user features (e.g., user's behavior history) and context features (e.g., operation time).

In addition, the statistics of the three datasets are summarized in Table~\ref{table:dataset}.

\subsubsection{Baselines}

We compare nine baseline models in our experiments, including LR~\cite{lr}, GBDT~\cite{gbdt}, FM~\cite{fm}, FFM~\cite{ffm}, CCPM~\cite{ccpm}, DeepFM~\cite{deepfm}, xDeepFM~\cite{xdeepfm}, IPNN and PIN~\cite{pin}. Wide \& Deep is not compared here because some state-of-the-art models (such as xDeepFM, DeepFM, PIN) have shown better performance in their publications. We use XGBoost~\cite{gbdt} and libFFM\footnote{https://github.com/guestwalk/libffm} as the implementation of GBDT and libFFM, respectively.  In our experiments, the other baseline models are implemented with Tensorflow\footnote{https://www.tensorflow.org/}.

\subsubsection{Evaluation Metrics} The evaluation metrics are \textbf{AUC} (Area Under ROC) and \textbf{log loss} (cross entropy).

\subsubsection{Parameter Settings}

Table~\ref{table:parameter} summerizes the hyper-parameters of each model.
For \textit{Criteo} and \textit{Avazu} Datasets, the hyper-parameters of baseline models are set to be the same as in~\cite{pin}. Notice that when conducting experiments on Criteo and Avazu, we observed that FGCNN uses more parameters in \emph{Deep Classifier} than other models. To make fair comparisons, we also conduct experiments to increase the parameters in MLPs of other deep models. However, all these models cannot achieve better performance than the original settings\footnote{Due to the limited pages, we do not show the experimental result in the paper.}. The reason could be the overfitting problem where such models simply use embedding of raw features for training but use a complex structure. On the other hand, since our model augments the feature space and enriches the input, more parameters in \emph{Deep Classifier} can boost the performance of our model.

In FGCNN model, \textit{new} is the number of kernels when generating new features. The number of generated features can be calculated as $\sum_{i=1}^{n_c} \#fields/2^i*new_i$.

\subsubsection{Significance Test} We repeat the experiments 10 times by changing the random seed for FGCNN and the best baseline model. The two-tailed pairwise t-test is performed to detect significant differences between FGCNN and the best baseline model. 
\begin{table*}[]
\caption{Compatibility Study of FGCNN}
\vspace{-3ex}
\label{table:integrete}
\resizebox{0.7\textwidth}{!}{
\begin{tabular}{c|ll|ll|ll}
\hline\hline
& \multicolumn{2}{c|}{Criteo} & \multicolumn{2}{c|}{Avazu} & \multicolumn{2}{c}{Huawei App Store} \\ \hline
& \multicolumn{1}{l}{AUC} & \multicolumn{1}{l|}{Log Loss} & \multicolumn{1}{l}{AUC} & \multicolumn{1}{l|}{Log Loss} & \multicolumn{1}{l}{AUC} & \multicolumn{1}{l}{Log Loss} \\ \hline
FM & 79.09\% & 0.5500 & 77.93\% & 0.3805 & 93.26\% & 0.1191 \\
FGCNN+FM & \textbf{79.67\%} & \textbf{0.5455} & \textbf{78.13\%} & \textbf{0.3794}  & \textbf{93.66}\% & \textbf{0.1165} \\ \hline
DNN & 79.87\% & 0.5428 & 78.30\% & 0.3778 & 93.85\% &0.1149 \\
FGCNN+DNN & \textbf{80.09\%} &\textbf{0.5402} & \textbf{78.55\%} & \textbf{0.3764} & \textbf{94.00\%} & \textbf{0.1139} \\ \hline
DeepFM & 79.91\% & 0.5423 & 78.36\% & 0.3777 & 93.91\% & 0.1145 \\
FGCNN+DeepFM & \textbf{79.94\%} & \textbf{0.5421}&\textbf{78.44\%} & \textbf{0.3771}& \textbf{93.93\%} & \textbf{0.1145} \\ \hline
IPNN & 80.13\% & 0.5399 & 78.68\% & 0.3757 & 93.95\% & 0.1143 \\
FGCNN+IPNN & \textbf{80.22\%} & \textbf{0.5388} & \textbf{78.83\%} & \textbf{0.3746} & \textbf{94.07\%} & \textbf{0.1134} \\ \hline
\hline

\end{tabular}
}
\vspace{-2ex}
\end{table*}

\subsection{Overall Performance (Q1)}
\footnotetext[11]{According to the experimental recordings, the results of PIN in ~\cite{pin} uses the trick of adaptive embedding size. Here, we use fixed embedding size for all the deep models.}

In this subsection, we compare the performance of different models on the test set. Table~\ref{table:overall} summarizes the overall performance of all the compared models on the three datasets, where the underlined numbers are the best results of the baseline models and bold numbers are the best results of all models.
We have the following observations:

Firstly, in majority of the cases, non-neural network models perform worse than neural network models. The reason is that deep neural network can learn complex feature interactions much better than the models where no feature interaction is modeled (i.e., LR), or feature interactions are modeled by simple inner product operations (i.e., FM and FFM).

Secondly,  FGCNN achieves the best performance among all the models on the three evaluated datasets. It is significantly better than the best baseline models with 0.05\%, 0.14\% and 0.13\% improvements in AUC (0.09\%, 0.24\% and 0.79\% in log loss) on Criteo, Avazu and Huawei App Store datasets, which demonstrates the effectiveness of FGCNN.  In fact, a small improvement in offline AUC is likely to lead to a significant increase in online CTR. As reported in~\cite{cheng2016wide}, compared with LR, Wide \& Deep improves offline AUC by 0.275\% and the improvement of online CTR is 3.9\%. The daily turnover of Huawei App Store is millions of dollars. Therefore, even a few lifts in CTR brings extra millions of dollars each year. 

Thirdly, with the help of the generated new features, FGCNN outperforms IPNN by 0.11\%, 0.19\% and 0.13\% in terms of AUC (0.2\%, 0.29\% and 0.79\% in terms of log loss) on Criteo, Avazu and Huawei App Store datasets. It demonstrates that the generated features are very useful and  they can effectively reduce the optimization difficulties of traditional DNNs thus leading to better performance.

Fourthly, CCPM, which applies CNN directly, achieves the worst performance among neural network models.  Moreover, CCPM performs worse than FFM on Criteo and Avazu datasets. It shows that directly using traditional CNN for CTR prediction task is inadvisable, as CNN is designed to generate neighbor patterns while the arrangement order of feature is usually no meaning in recommendation scenarios. However, in FGCNN, we leverage the strength of CNN to extract local patterns while complementing it with Recombination Layer to extract global feature interactions and generate new features. Therefore, better performance is achieved.
\subsection{Compatibility of FGCNN with Different Models (Q2)}

As stated in Section ~\ref{sec:clf}, \emph{Feature Generation}  can augment the original feature space and \emph{Deep Classifier} of FGCNN can adopt any advanced deep neural networks. Therefore, we select several models  as \emph{Deep Classifeir} to verify the utility of \emph{Feature Generation}, including non-deep models (FM), deep learning models (DNN, DeepFM, IPNN).

Table~\ref{table:integrete} summarizes the performance. We have the following observations: Firstly, with the help of the generated new features, the performance of all models are improved, which demonstrates the effectiveness of the generated features and shows the compatibility of FGCNN. Secondly, we observe that when only using raw features, DeepFM always outperforms DNN. But when using the augmented features, FGCNN+DNN outperforms FGCNN+DeepFM. The possible reason is that DeepFM sums up the inner product of input features to the last MLP layer which may cause contradictory gradient updates (compared with MLP) on embeddings.
This could be one of the reasons why IPNN (feeding the product into MLP) outperforms DeepFM in all datasets.

In a word, the results show that our FGCNN model can be viewed as a general framework to enhance the existing neural networks by generating new features automatically.

\subsection{Effectiveness of FGCNN Variants(Q3)}
\begin{table*}[t]
\vspace{-2ex}
\caption{Performance of Different FGCNN Variants}
\vspace{-3ex}
\label{table:remove}
\resizebox{0.8\textwidth}{!}{
\begin{tabular}{l|ll|ll|ll}
\hline\hline
       & \multicolumn{2}{c|}{Criteo} & \multicolumn{2}{c|}{Avazu} & \multicolumn{2}{c}{Huawei App Store} \\ \hline
Method & AUC          & Log Loss     & AUC         & Log Loss     & AUC          & Log Loss     \\ \hline
FGCNN  & \textbf{80.22\%}      & \textbf{0.5388}      & \textbf{78.83\%}     & \textbf{0.3746}       & \textbf{94.07\%}      & \textbf{0.1134}       \\
Removing Raw Features   & 80.21\%      & 0.5390       & 78.66\%     & 0.3757        & 94.01\%      &0.1138       \\
Removing New Features   & 80.13\%      & 0.5399       & 78.68\%     & 0.3757       &  93.95\%      &  0.1143      \\
Applying MLP for Feature Generation   & 80.12\%      & 0.5402       & 78.58\%     & 0.3761       & 94.04\%             &0.1135              \\
Removing Recombination Layer    & 80.11\%      & 0.5403       & 78.74\%     & 0.3753       & 94.04\%             & 0.1135             \\
\hline\hline
\end{tabular}
}
\vspace{-2ex}
\end{table*}

 We conduct experiments to study how each component in FGCNN contributes to the final performance. Each variant is generated by removing or replacing some components in FGCNN, which is described as follows:

\begin{itemize}
\item \textbf{Removing Raw Features}: In this variant, raw features are not input into \textit{Deep Classifier} and only the  generated new  features are fed to \emph{Deep Classifier}.
 \item \textbf{Removing New Features}: This variant removes \emph{Feature Generation}. Actually, it is equivalent to IPNN.
 \item \textbf{Applying MLP for \textit{Feature Generation}}: \emph{Feature Generation} is replaced by MLP which takes the neurons in each layer as new features. This variant uses the same hidden layers and generates the same number of features in each layer as FGCNN.
 \item \textbf{Removing Recombination Layer}: This variant is to evaluate how \emph{Recombination Layer} complements CNN to capture global feature interactions.  The \emph{Recombination Layer} is removed from \emph{Feature Generation}   so that the output of pooling layer serves as new features directly. The number of generated new features in each layer keeps the same as FGCNN.
\end{itemize}

As shown in  Table~\ref{table:remove}, removing any component in FGCNN leads to a drop in performance. We have the following observations:

Firstly, FGCNN with raw features alone or with new generated features alone performs worse than the FGCNN with both of them. This result demonstrates that the generated features are good supplementaries to the original features, which are both crucial.

Secondly, the performance decrease of \textit{Appling MLP for Feature Generation}, compared to FGCNN, shows the ineffectiveness of MLP to identify the sparse but important feature combinations from a large number of parameters. CNN simplifies the learning difficulties by using the shared convolutional kernels, which has much fewer parameters to get the desired combinations. Moreover,  MLP recombines the neighbor feature interactions, which is extracted by CNN, to generate global feature interactions.

Thirdly, removing the Recombination Layer will limit the generated features as neighbor feature interactions. Since the arrangement order of raw features has no actual meanings in the CTR prediction task,  the restriction can lead to losing important non-neighbor feature interactions thus resulting in worse performance.


\subsection{Hyper-parameter Investigation (Q4)}

Our FGCNN model has several key hyper-parameters, i.e., the height of convolutional kernels, number of convolutional kernels, number of convolutional layers, and the number of kernels for generating new features. In this subsection, to study the impact of these hyper-parameters, we investigate how FGCNN model works by changing one hyper-parameter while fixing the others on Criteo and Huawei App Store datasets.

\subsubsection{Height of Convolutional Kernels}
 
\begin{figure}[t]
 \centering
 \includegraphics[width=0.45\textwidth,height=9cm]{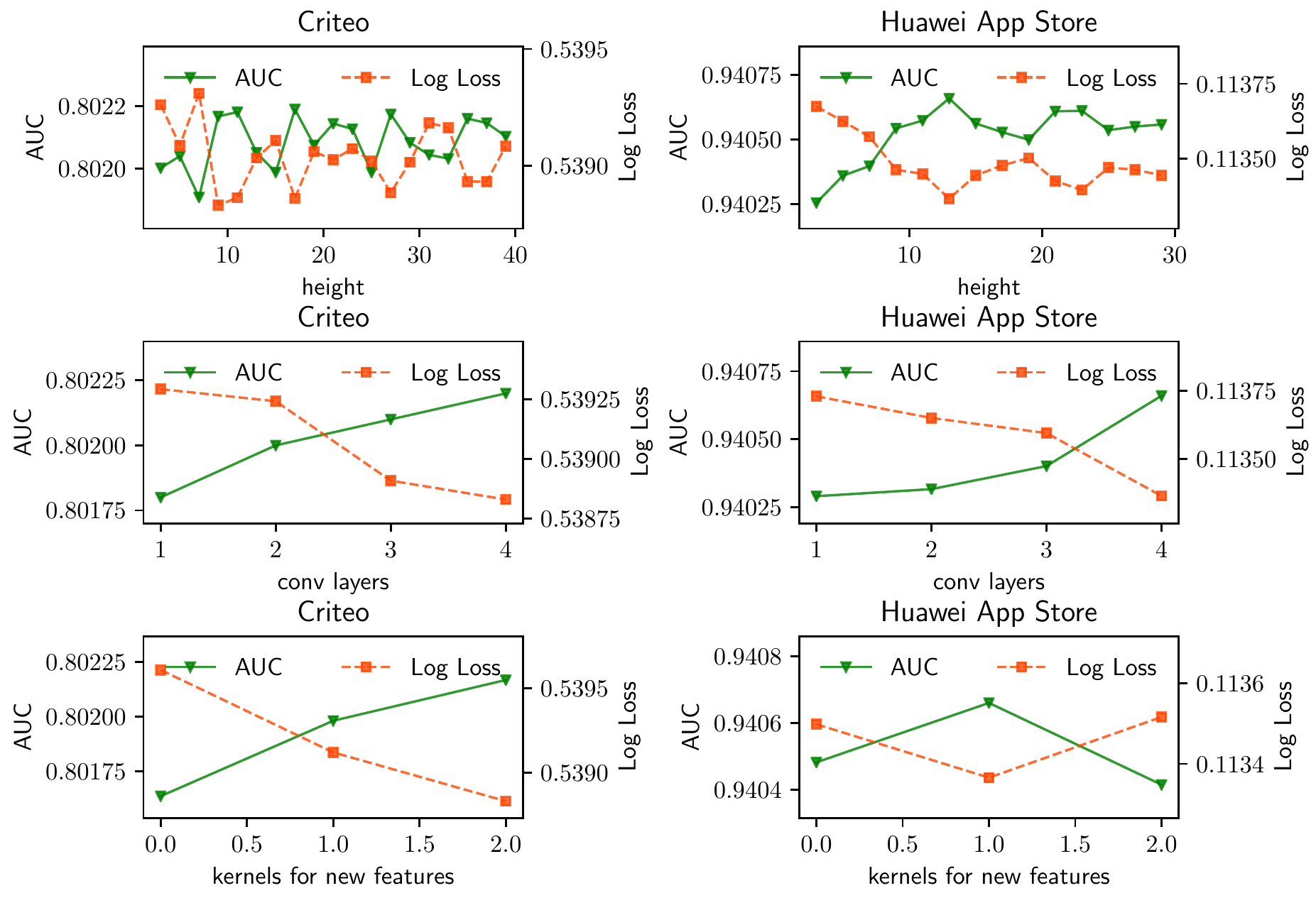}
 \vspace{-4ex}
 \caption{Parameter study of height of convolutional kernel, number of convolutional layers and number of  kernels for new features (from top to bottom)}
 \vspace{-3ex}
 \label{fig:param_all}
\end{figure}

The height of convolutional kernels controls the perception range of convolutional layers. The larger the height is, the more features are involved in the neighbor patterns, but more parameters need to be optimized. To investigate its impact, we increase the height from 2 to the number of fields of a dataset. As shown in the top of Figure~\ref{fig:param_all},
the performance generally ascends first and then descends as the height of convolutional kernels increases\footnote{Due to that \textit{Feature Generation} and \textit{Deep Classifier} interrelate with each other, the curve has some fluctuations.}.

The results show that as more features are involved in the convolutional kernels, higher-order feature interactions can be learned so that the performance increases. However, due to that useful feature interactions are usually sparse, larger heights can cause more difficulties to learn them effectively, leading to a decrease in performance. This observation is consistent with the finding in Section 3.4, i.e., the performance decreases in Appling MLP for Feature Generation.

\subsubsection{Number of Convolutional Layers}
As shown in  the middle of  Figure~\ref{fig:param_all}, as the number of convolutional layers increases, the performance of FGCNN is improved. Notice that more layers usually lead to higher-order feature interactions. Therefore, the result also shows the effectiveness of high-order feature interactions. 

\subsubsection{Number of Kernels for Generating New Features}
We study how the number of generated features affects the performance of FGCNN. We use the same number of kernels for new features in different \emph{Recombination Layers}.  As can be observed in the bottom of Figure~\ref{fig:param_all}, the performance is gradually improved with more features generated. The results verify our research idea that it is useful to identify the sparse but important feature interactions first, which can effectively reduce the learning difficulties of DNNs. However, the useful feature interactions can be sparse and limited. If too many  features are generated, the extra new features are noisy which will increase the learning difficulties of MLP,  leading to the decrease in the performance.

\subsection{Effect of Shuffling Order of Raw Features (Q5)}
\begin{figure}[h]
 \centering
\vspace{-2ex}
 \includegraphics[width=0.45\textwidth]{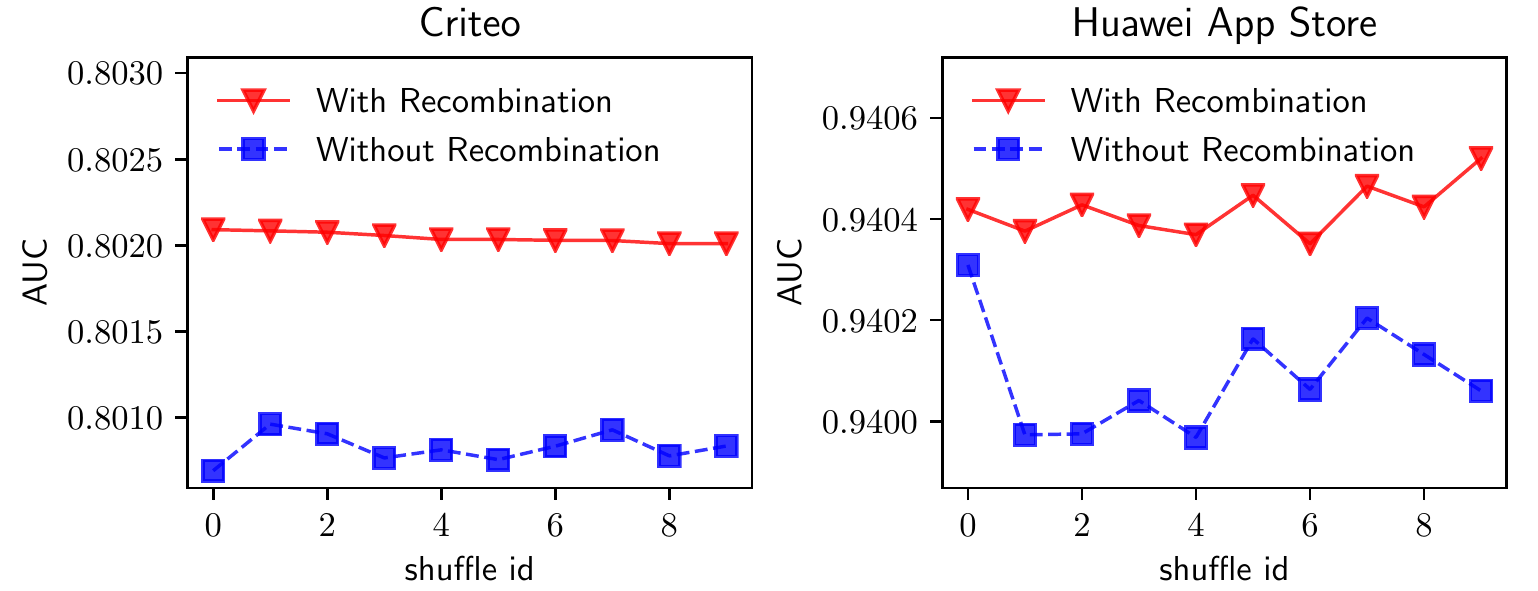}
 \vspace{-3ex}
 \caption{Shuffling the order of raw features}\label{fig:shuffle}
 \vspace{-3ex}
 \label{shuffle}
\end{figure}

As mentioned before, CNN is designed to capture local neighbor feature patterns so that it is sensitive to the arrangement order of raw features. In our FGCNN model, the design of Recombination Layer is to learn global feature interactions based on CNN's extracted local patterns. Intuitively, our model should have more stable performance than traditional CNN's structure if the order of raw features is shuffled. Therefore,  to verify it, we compare the performance of two cases: with/without Recombination Layer. The arrangement order of raw features is shuffled many times at random where the two compared cases are performed for the same shuffled arrangement order.

As shown in  Figure~\ref{fig:shuffle}, the case with Recombination Layer achieves better and more stable performance than that of without Recombination Layer. It demonstrates that with the help of Recombination Layer, FGCNN can greatly reduce the side effects of changing the arrangement order of raw features, which also demonstrates the robustness of our model.

%% file: related_work.tex
\vspace{-1ex}
\section{Related Work}
Click-Through Rate Prediction is normally formulated as a binary classification problem~\cite{Richardson2007Predicting,Chen2016Deep}. In this section, we will introduce two important categories of models in Click-Through Rate predictions, namely \textbf{shallow models} and \textbf{deep learning models}. 

\subsection{Shallow Models for CTR Prediction}
Due to the robustness and efficiency,  Logistic Regression (LR) models~\cite{Ren2016User,Lee2012Estimating}, such as FTRL~\cite{ftrl} are widely used in CTR prediction. To learn feature interactions, a common practice is to manually design pairwise feature interactions in its feature space~\cite{deepfm2,Sun2017Practical}. Poly-2~\cite{poly2} models all pairwise feature interactions to avoid feature engineering. Factorization Machine (FM)~\cite{fm} introduces  low-dimensional vectors for each feature and models feature interactions through inner product of feature vectors. FM improves the ability of modelling feature interactions when data is sparse.  FFM~\cite{ffm} enables each feature to have multiple latent vectors to perform interactions with features  from different fields. LR, Poly-2 and FM variants are widely used in CTR prediction in industry. 

\subsection{Deep Learning for CTR Prediction}
Deep learning has achieved great success in different areas, such as computer vision~\cite{Kaiming2015Deep,Szegedy2016Inception,Huang2016Densely,Xie2017Aggregated}, natural language processing~\cite{Bahdanau2014Neural,Bordes2016Learning,Radford2017Learning,Vaswani2017Attention} etc. In order to leverage deep Learning for CTR prediction, several models are proposed~\cite{crossnet,deepsurvey}. FNN~\cite{fnn} is proposed in ~\cite{fnn}, which uses FM to pre-train the embedding of raw features and then feeds the embeddings to several fully connected layers. Some models adopt DNN to improve FM, such as Attentional FM~\cite{afm}, Neural FM~\cite{nfm}. 

Wide \& Deep Learning~\cite{cheng2016wide} jointly trains a wide model and a deep model where the wide model leverages the effectiveness of feature engineering and the deep model learns implicit feature interactions. Despite the usefulness of wide component, feature engineering is expensive and requires expertise. 
To avoid feature engineering, DeepFM~\cite{deepfm} introduces FM layer (order-2 feature interaction) as a wide component and uses a deep component to learn implicit feature interactions. Different from DeepFM, IPNN~\cite{pin} (also known as PNN in ~\cite{pnn}) feeds both the result of FM layer and embeddings of raw features into MLP   and results in comparable performance. Rather than using inner product to model pairwise feature interactions as DeepFM and IPNN, PIN~\cite{pin} uses a Micro-Network  to learn complex feature interaction for each feature pair. xDeepFM~\cite{xdeepfm} proposes a novel Compressed Interaction Network (CIN) to explicitly generate feature interactions at the vector-wise level.

There are several models which use CNN for CTR Prediction. 
CCPM~\cite{ccpm} applies multiple convolutional layers to explore neighbor feature dependencies. CCPM performs convolutions on the neighbor fields in a certain alignment. Due to that the order of features has no actual meaning in CTR predictions, CCPM can only learn limited feature interactions between neighbor features. In ~\cite{cnntencent}, it is shown that features' arrangement order has a great impact on the final performance of CNN based models. Therefore, the authors propose to generate a set of suitable feature sequences to provide different local information for convolutional layers. However, the key weakness of CNN is not solved. 

In this paper, we propose FGCNN model, which splits the CTR prediction task into two stages: \textit{Feature Generation} and \textit{Deep Classifier}.  \textit{Feature Generation} augments the original feature space by generating new features while most state-of-the-art models can be adopted in \textit{Deep Classifier} to learn and prediction based on the augmented feature space. Different from traditional CNN models for CTR Prediction~\cite{ccpm,cnntencent}, FGCNN can leverage the strength of CNN to extract local information and it greatly alleviates the weakness of CNN by introducing the \textit{Recombination Layer} to recombine information from different local patterns learned by CNN to generating new features.